# Inclusive Kitchen Design for Older Adults: Generative AI Visualizations to Support Mild Cognitive Impairment

Ibrahim Bilau, Georgia Institute of Technology, USA
Nicole Li, Georgia Institute of Technology, USA
Terrence Malayvong, Georgia Institute of Technology, USA
Eunhwa Yang, PhD, Georgia Institute of Technology, USA

## Abstract

Mild Cognitive Impairment (MCI) affects 15–20% of adults aged 65 and older, often making kitchen navigation and independent living difficult, particularly in lower-income communities with limited access to professional design help. This study created an AI system that converts standard kitchen photos into MCI-friendly designs using the Home Design Guidelines (HDG). Stable Diffusion models, enhanced with DreamBooth LoRA and ControlNet, were trained on 100 kitchen images to produce realistic visualizations with open layouts, transparent cabinetry, better lighting, non-slip flooring, and less clutter. The models achieved moderate to high semantic alignment (normalized CLIP scores 0.69–0.79) and improved visual realism (GIQA scores 0.45–0.65). In a survey of 33 participants (51.5% caregivers, 36.4% older adults with MCI), the AI-modified kitchens were strongly preferred as more cognitively friendly (87.4% of 198 choices, $p < .001$). Participants reported high confidence in their kitchen choice selections ($M = 5.92/7$) and found the visualizations very helpful for home modifications ($M = 6.27/7$). Thematic analysis emphasized improved visibility, lower cognitive load, and greater independence. Overall, this AI tool provides a low-cost, scalable way for older adults and caregivers to visualize and implement DIY kitchen changes, supporting aging in place and resilience for those with MCI.

## Introduction

Mild Cognitive Impairment (MCI) involves a noticeable decline in cognitive functions, such as memory, attention, and visuospatial abilities, beyond typical age-related changes but not severe enough to qualify as dementia (Petersen et al., 2018). MCI affects an estimated 15–20% of adults aged 65 and older worldwide (Bai et al., 2022). Older adults with MCI often struggle with instrumental activities of daily living (IADLs), like meal preparation, financial management, and household maintenance, which are critical for maintaining independence (Jekel et al., 2015). These challenges are particularly evident in high-risk areas of the home, such as the kitchen, bathroom, and living spaces. The kitchen stands out as a vital space, linked to both activities of daily living (ADLs), such as eating, and IADLs, such as cooking, making it a key focus for interventions to support independent living (Lawton & Brody, 1969). Recent research further shows that specific home design factors significantly influence how older adults with MCI perform IADLs, underscoring the importance of evidence-based environmental modifications (Machry et al., 2025; Wahl et al., 2012).

Cognitive and functional declines associated with MCI often require modifications to the home environment to enhance safety, usability, and comfort. For example, visuospatial deficits can impair navigation and object recognition, while memory issues may make it hard to recall task sequences (Farias et al., 2006). Recent studies confirm that features such as open shelving significantly reduce cognitive load and improve task performance for older adults living with MCI (Bilau et al., 2025; Johansson et al., 2011). Universal design principles, which emphasize creating environments usable by people of all abilities, provide the overarching framework for such modifications (Sanford, 2012; Steinfeld & Maisel, 2012). To address these needs, Yang et al. (2023) developed the Home Design Guidelines (HDG), a comprehensive resource offering evidence-based design strategies and recommendations tailored for individuals with MCI and their caregivers. The HDG suggests modifications ranging from simple do-it-yourself changes, such as decluttering spaces, changing colors, adding labels, or improving lighting, to extensive renovations, such as reconfiguring layouts or integrating smart home technologies.

Despite its value, the HDG faces barriers to adoption, primarily due to the cognitive limitations of individuals with MCI. Research shows that people with MCI struggle to process complex textual information and benefit more from visual aids (Albert et al., 2011). Visual representations are often more intuitive, aiding comprehension and decision-making (Bourgeois, 1992). However, translating the HDG's textual recommendations into personalized visual designs typically requires professional architects or interior designers, which can be costly and time-consuming. This financial and logistical burden highlights the need for an innovative, accessible solution to bridge the gap between the HDG and its practical application.

Recent advancements in artificial intelligence (AI) offer a promising way forward. AI-driven image generation technologies, such as stable diffusion models, excel at creating realistic, contextually relevant images from text prompts or existing visuals (Rombach et al., 2022). By harnessing these tools, it may be possible to transform the HDG's recommendations into MCI-friendly visual renderings of home environments, reducing the cognitive strain on individuals with MCI and their caregivers. This study explores the use of stable diffusion models, enhanced by techniques like DreamBooth LoRA and ControlNet, to generate image-to-image transformations of existing kitchen interiors into MCI-friendly designs.

Stable diffusion is a latent diffusion model that produces high-quality images from text prompts or modifies existing images based on specific constraints (Rombach et al., 2022). Other models, such as DALL·E 2 (Ramesh et al., 2022) and VQ-VAE-2 (Razavi et al., 2019), were considered but deemed less suitable. DALL·E 2 lacks precision for fine-grained image modifications, and VQ-VAE-2 struggles with detailed architectural rendering. Within the stable diffusion framework, DreamBooth LoRA (Low-Rank Adaptation) supports fine-tuning pre-trained models on specific datasets for personalized image generation (Hu et al., 2021). ControlNet adds spatial control, ensuring accurate layout and object placement, which is ideal for architectural applications (Zhang et al., 2023; Hattori et al., 2024). These tools were chosen for their balance of flexibility, precision, and computational efficiency.

Given the wide range of possible home modifications, this study focuses on the kitchen, a critical space for independent living due to its role in ADLs and IADLs (Lawton & Brody, 1969). The research questions guiding this work are:

**RQ1.** Can an AI model, specifically stable diffusion with DreamBooth LoRA and ControlNet, effectively generate MCI-friendly kitchen designs from existing kitchen images?

**RQ2.** Do the generated images align with HDG recommendations for usability, safety, and accessibility when evaluated through semantic alignment with prompts and assessments of visual quality and realism?

**RQ3.** To what extent do older adults with MCI and their caregivers perceive the AI-generated designs as more cognitively friendly, realistic, and helpful for planning home modifications?

To answer these questions, the study adopts a multi-phase approach: (1) data collection using HDG-derived prompts, (2) model training, (3) technical evaluation with CLIP and GIQA metrics, and (4) user validation through a survey with older adults with MCI and their caregivers.

This research aims to advance the intersection of AI and assistive technologies by offering a scalable, cost-effective solution to enhance independence and quality of life for individuals with MCI.

## Methods

This study employed a mixed-methods design to develop and evaluate an AI-driven system capable of transforming standard kitchen images into MCI-friendly designs aligned with the evidence-based HDG. The methodology consisted of two main components: (1) development and technical evaluation of the generative AI model and (2) user validation through an online survey. All procedures were reviewed and approved by the Georgia Institute of Technology Institutional Review Board (IRB) as exempt research involving minimal risk. Electronic informed consent was obtained from all participants prior to data collection.

**AI Model Development**

A Python-based tool was developed to collect high-quality images of universal-design-friendly kitchens from Unsplash, a platform providing openly licensed stock photography. The Unsplash API was utilized for its reliability and simplicity. Approximately 100 images were gathered along with associated metadata describing design features. The permissive licensing of Unsplash ensured full legal compliance for research purposes. All images were preprocessed by resizing to 512 × 512 pixels and normalizing pixel values, then manually annotated with HDG-derived text prompts, such as "open shelving," "transparent cabinetry," "non-slip flooring," and "under-cabinet lighting." The dataset was partitioned into training (80%), validation (10%), and test (10%) sets.

*Phase 2: Model Training*

Training was conducted on a single NVIDIA GeForce RTX 3080 GPU. Four progressive model variants (M1–M4) were developed using Stable Diffusion enhanced with DreamBooth

Low-Rank Adaptation (LoRA) and ControlNet. Model M1 served as the baseline and was trained for 10,000 steps using HDG-focused prompts emphasizing open layouts and transparent cabinetry. Models, M2-M4, underwent additional fine-tuning for 15,000–20,000 steps, incorporating ControlNet's Canny edge and depth conditioning to improve spatial coherence. A learning rate of $1 \times 10^{-4}$, a batch size of 4, and a 50% random replacement of prompts with empty strings were used to enhance robustness, following established practices (Zhang et al., 2023). Negative prompts such as "clutter" and "dark" were consistently used to prioritize accessibility-focused outputs.

*Phase 3: Technical Evaluation*

Ten held-out test images were processed through each model. Generated outputs were evaluated using two established metrics: CLIP Score for semantic alignment between HDG prompts and the resulting images (Radford et al., 2021), and Generated Image Quality Assessment (GIQA) for visual quality and realism (Gu et al., 2020). These metrics were selected over alternatives such as Fréchet Inception Distance (Kynkäänniemi et al., 2019; Sajjadi et al., 2018) because they better capture functional design alignment rather than mere diversity. CLIP and GIQA scores were normalized to a 0–1 scale for interpretability.

*Phase 4: User Validation Survey*

To assess real-world acceptability, an anonymous online survey was administered via Qualtrics. Participants were recruited exclusively from members and alumni of a cognitive empowerment program at Emory University, a community-based initiative designed specifically for older adults living with MCI. The survey presented six randomized pairs of images (standard kitchen versus the corresponding AI-modified version). For each pair, participants selected which kitchen appeared more cognitively friendly (easier, safer, and more usable for someone with memory or attention difficulties), rated their confidence in the choice (1–7 scale), and indicated how helpful such visualizations would be for planning home modifications (1–7 scale). Open-ended explanations and general reflections were also collected. Thirty-three participants completed the survey: 17 caregivers or care partners (51.5%), 12 older adults with MCI (36.4%), 2 older adults without MCI (6.1%), and 2 others (6.1%).

**Figure 1**
*Four-Step Methodological Framework: Data Collection, Model Training, Technical Evaluation, and User Validation*

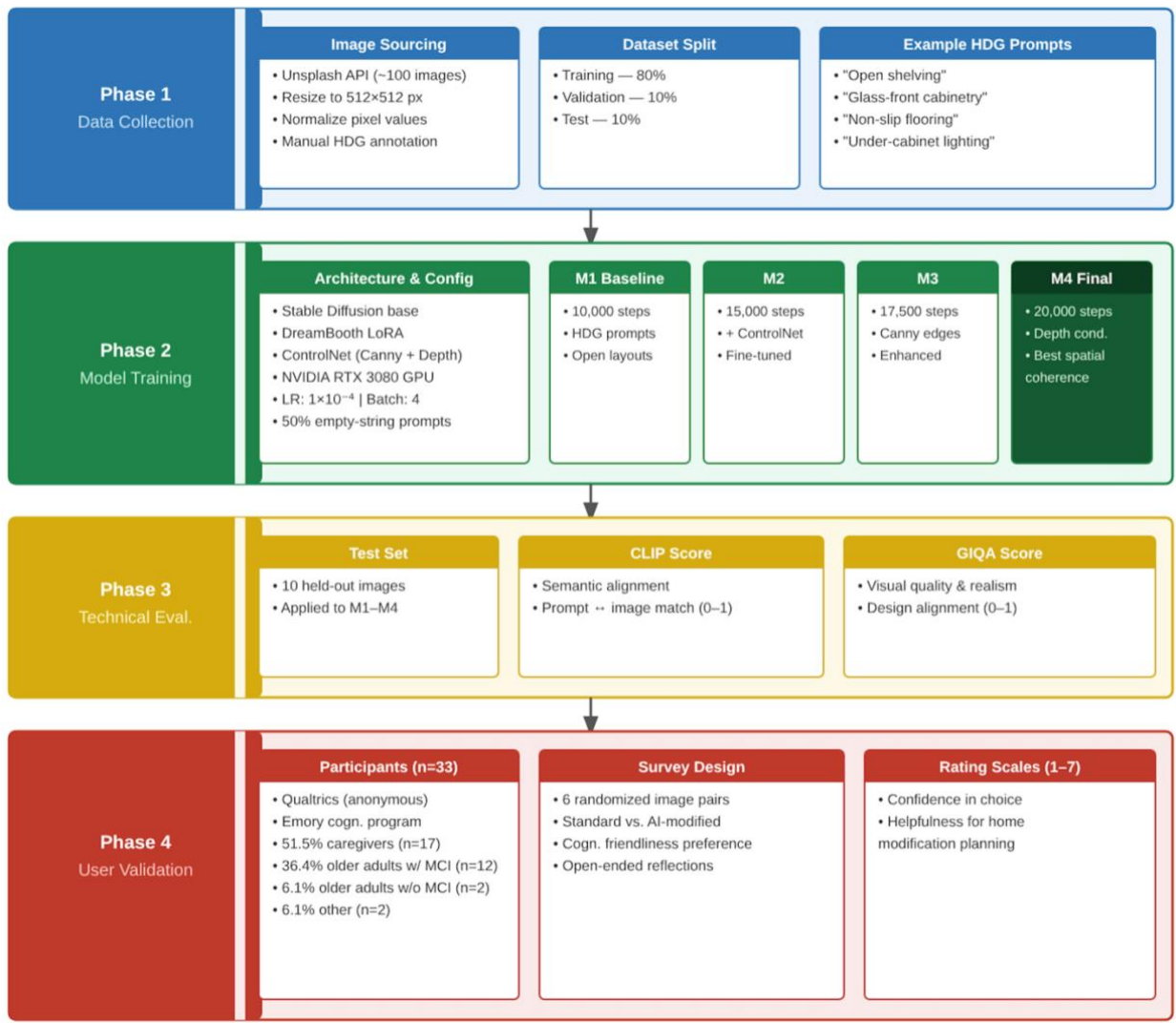

**Results**

This study evaluated the AI system through two complementary approaches: objective technical metrics on the generative models and direct user validation via a survey. Together, these analyses addressed all three research questions by confirming technical alignment with the HDG and strong real-world acceptability among older adults and caregivers.

**AI Model Performance**

Four progressive model variants (M1–M4) plus a baseline (M0) were tested on 10 held-out kitchen images. Semantic alignment with HDG prompts was measured using normalized CLIP scores (0–1 scale). Mean normalized scores showed a gradual decline from M1 (0.79) to M4 (0.69), while M0 averaged 0.67. Individual image scores ranged from approximately 0.16 to 1.00 (see Figure 2 and Table 1). Although raw CLIP means ranged from 30.93 for M1 to 29.57 for M4, this modest reduction reflects progressive specialization rather than degradation. Later models incorporated MCI-specific features such as transparent cabinetry, assistive elements, and reduced clutter, which are underrepresented in CLIP's general training data (Schuhmann et al., 2022), resulting in deliberate divergence from generic kitchen imagery rather than a loss of semantic quality.

Visual realism and coherence were assessed using GIQA metrics. Normalized GMM scores

(global realism) were highest for M1 at approximately 0.94 and declined progressively to approximately 0.58 for M4. Individual values ranged from approximately 0.08 to 1.00 (see Figure 3 and Table 2). Normalized KNN scores (local feature coherence) followed a similar pattern, with M1 at approximately 0.95 decreasing to approximately 0.32 for M4 (see Figure 4 and Table 3).

These patterns indicate that later models learned a coherent MCI-friendly design distribution distinct from standard kitchens. Average scores across metrics are summarized in Figure 5, and representative visual comparisons appear in Figure 6. Expert review confirmed that M2–M4 preserved spatial layout while clearly integrating open shelving, improved lighting, and non-slip flooring.

Although later models (M2–M4) produced outputs that were increasingly distinctive of MCI-friendly design, M1 consistently showed the strongest overall performance across CLIP, GMM, and KNN metrics, indicating that it was best across semantic alignment and visual realism, and was therefore selected as the final model. The six image pairs used in the user survey were generated with M1 applied to standard American kitchen photographs. Collectively, the metrics confirm that the models effectively generated HDG-aligned designs (RQ1 and RQ2), with M1 providing the optimal balance for real-world application.

**Figure 2**
*Normalized CLIP similarity scores per image*

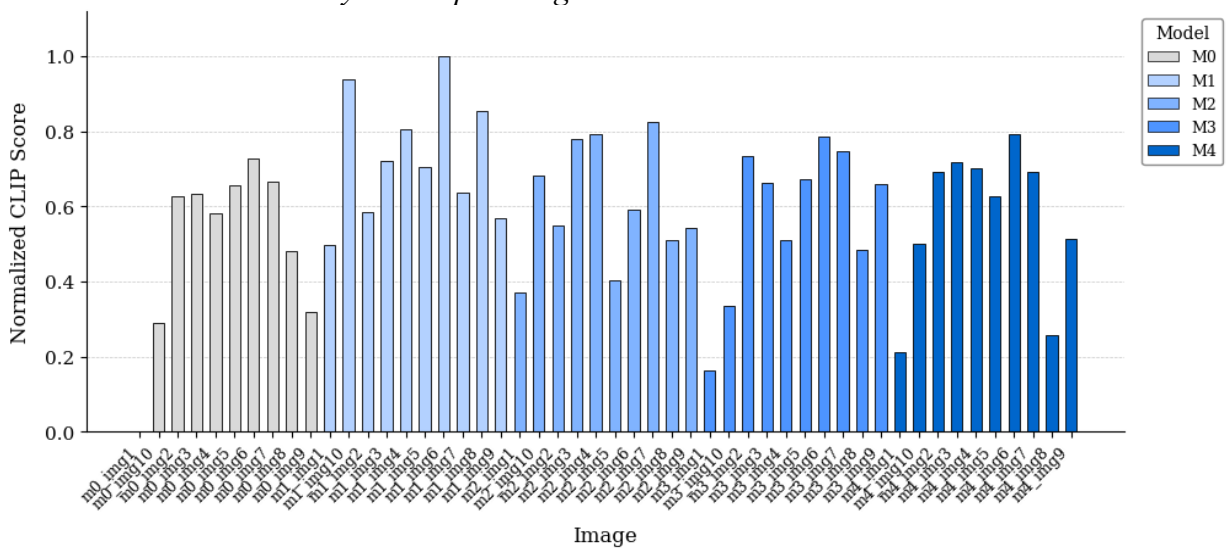

**Table 1**

*CLIP Scores Per Model. % Quality Relative to Best Model (M1-M4). Vs Input = % Change Relative to M0 Baseline.*

| Model | Mean | % Quality | vs Input | Rank |
|---|---|---|---|---|
| M0 | 28.9447 | — | — | — |
| M1 | 30.9256 | 100.0% | +6.8% | #1 |
| M2 | 29.8510 | 96.5% | +3.1% | #2 |
| M3 | 29.5998 | 95.7% | +2.3% | #3 |
| M4 | 29.5652 | 95.6% | +2.1% | #4 |

**Figure 3**

*Normalized GIQA GMM Scores Per Image*

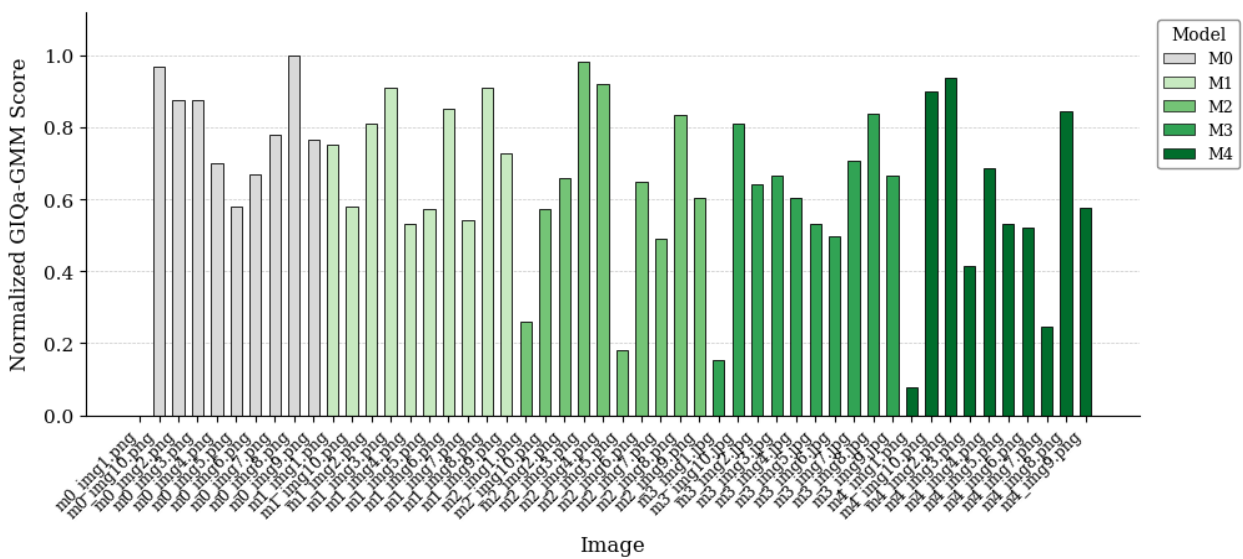

**Table 2**

*Raw GIQA-GMM Log-Likelihood Scores Per Model. % Quality Relative to Best Model (M1-M4). Vs Input = % Change Relative to M0 Baseline.*

| Model | Mean | % Quality | vs Input | Rank |
|---|---|---|---|---|
| M0 | -7,640,876.0659 | — | — | — |
| M1 | -7,668,629.0891 | 100.0% | -0.4% | #1 |
| M2 | -9,083,346.2760 | 84.4% | -18.9% | #2 |
| M3 | -9,135,986.5661 | 83.9% | -19.6% | #3 |

| M4 | -9,643,285.8763 | 79.5% | -26.2% | #4 |

**Figure 4**

*Normalized GIQA KNN Scores Per Image*

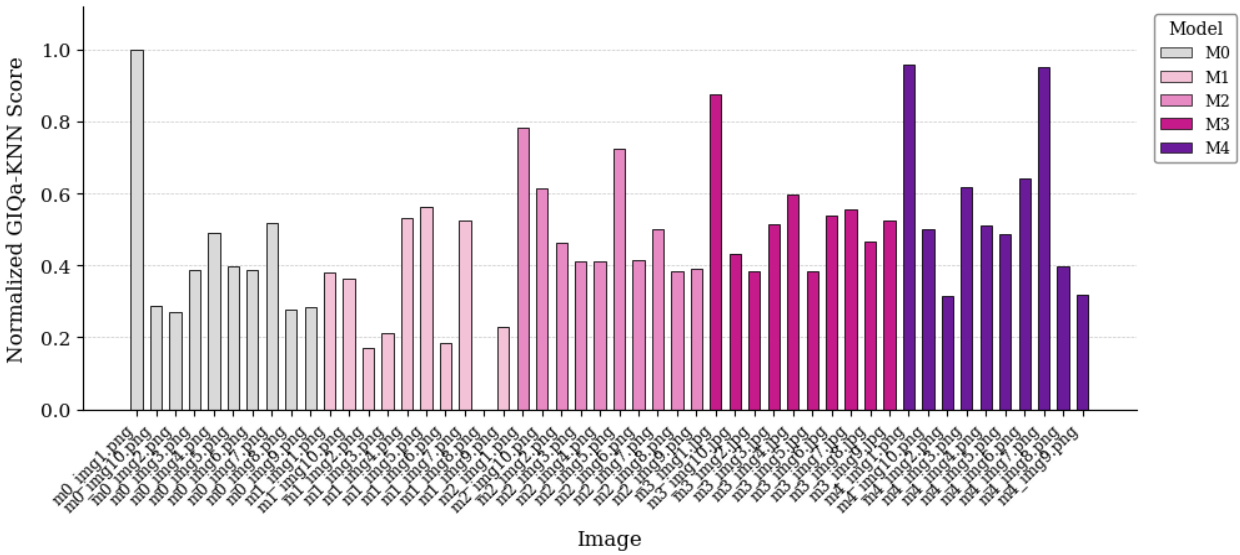

**Table 3**

*Raw GIQA-KNN Log-Likelihood Scores Per Model. % Quality Relative to Best Model (M1-M4). Vs Input = % Change Relative to M0 Baseline.*

| Model | Mean | % Quality | vs Input | Rank |
| --- | --- | --- | --- | --- |
| M0 | 9.3586 | — | — | — |
| M1 | 8.8503 | 100.0% | +5.4% | #1 |
| M2 | 9.7139 | 91.1% | -3.8% | #2 |
| M3 | 9.7910 | 90.4% | -4.6% | #3 |
| M4 | 9.9784 | 88.7% | -6.6% | #4 |

**Figure 5**

*Average Normalized CLIP, GIQA-GMM, and GIQA-KNN Scores Per Model*

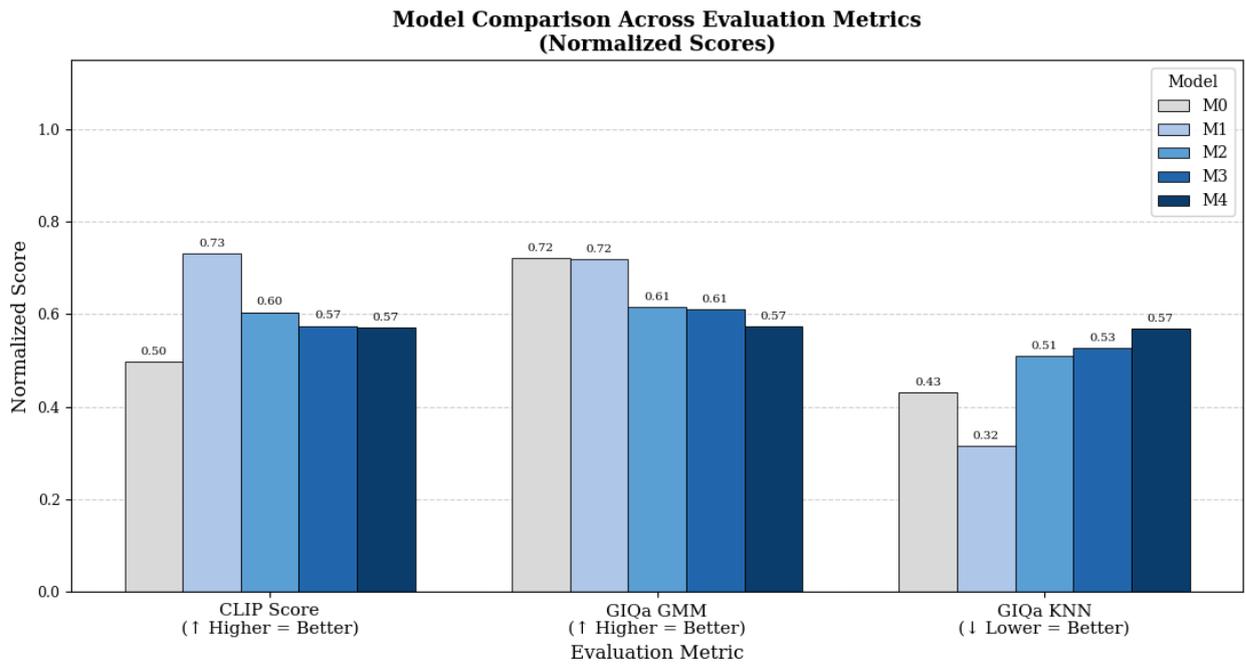

**Figure 6**
*Visual and Quantitative Model Comparison Per Model*

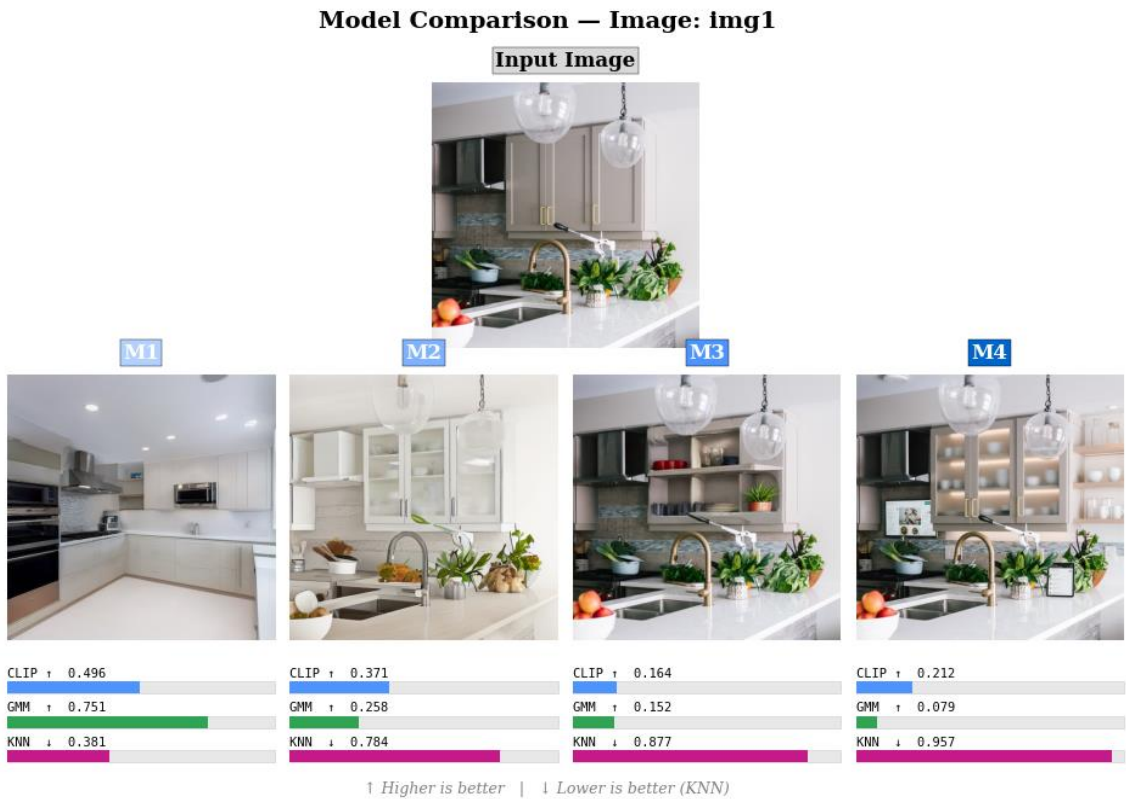

**User Validation Results**

To assess real-world acceptability, an anonymous online survey was administered via the Qualtrics platform. Participants were recruited exclusively from members and alumni of a cognitive empowerment program at Emory University. A total of 33 participants completed the survey. The sample was predominantly older adults (51.5%, n = 17, aged 70–79; 33.3%, n = 11, aged 80 or older; 12.1%, n = 4, aged 60–69; 3.0%, n = 1, aged 50–59), with 51.5% (n = 17) identifying as caregivers or care partners, 36.4% (n = 12) as older adults with MCI, 6.1% (n = 2) as older adults without MCI, and 6.1% (n = 2) as other. Most were highly engaged in kitchen activities (60.6% cooking daily, 24.2% frequently), aligning with the target population.

Participants viewed the six randomized image pairs generated by the selected M1 model (see Figure 7). For each pair, they selected which kitchen appeared more cognitively friendly (easier, safer, and more usable for memory or attention difficulties), rated their confidence in the choice (1 to 7 scale), and indicated how helpful such visualizations would be for planning home modifications (1 to 7 scale). Open-ended explanations and general reflections were also collected.

Of 198 total choices, 173 (87.4%) favored the AI-modified kitchens (p < .001 for each pair and overall; 95% CI [82.8%, 100%]).

**Table 5**
*Preference for Optimized Kitchens by Image Pair (N = 33)*

| Image Pair | Percent Choosing Optimized | Successes | p-value (binomial one-sided) | 95% CI |
| --- | --- | --- | --- | --- |
| Pair 1 | 100 % | 33 | < .001 | [89.4, 100] |
| Pair 2 | 79 % | 26 | < .001 | [60.3, 91.3] |
| Pair 3 | 88 % | 29 | < .001 | [71.8, 96.6] |
| Pair 4 | 91 % | 30 | < .001 | [75.7, 98.1] |
| Pair 5 | 82 % | 27 | < .001 | [64.5, 93.0] |
| Pair 6 | 85% | 28 | < .001 | [67.5, 95.2] |
| Overall | 87.4 % | 173/198 | < .001 | [82.8, 100] |

Confidence in choices was high (M = 5.92 out of 7 across pairs, SD ≈ 1.1), significantly above the neutral midpoint of 4 (one-sample t-test, p < .0001). Perceived helpfulness of AI-generated visualizations for planning home modifications (Q7, 1 to 7 scale) was also strongly positive (M = 5.7/7), with 72.7% (n = 24) rating them as "Helpful" (6) or "Extremely Helpful" (7). Only two respondents (6.1%) rated the visualizations negatively. Subgroup analyses showed no statistically significant differences.

Thematic analysis of open-ended responses (Braun & Clarke, 2006) identified two primary themes.

The dominant positive theme, visibility and reduced cognitive load, which were mentioned in more than 80% of responses, centered on open shelving or transparent cabinetry *("Can see everything without opening doors and guessing")*, minimal clutter, enhanced lighting (under-cabinet or kick-plate), and visual cues (labels, fridge notes, recipe tablets). Participants noted these features minimized memory search and distraction, directly supporting safer, more independent meal preparation and activities in the kitchen.

A secondary theme, practical concerns and realism (approximately 35 to 40% of responses),

highlighted potential drawbacks of open shelving (reaching high shelves, risk of disorganization or dust in daily use), and loss of counter space. Many appreciated the designs for MCI but questioned long-term maintainability *("I love it for MCI, but I would not keep open shelves perfectly organized")*. Realism ratings were more mixed than preference scores, with several noting the optimized versions appeared "too perfect or sterile" for real homes.

**Figure 7**

*Original and M1-Generated Cognitive-Friendly Versions of The Six Kitchen Images Used in the User Validation Survey.*

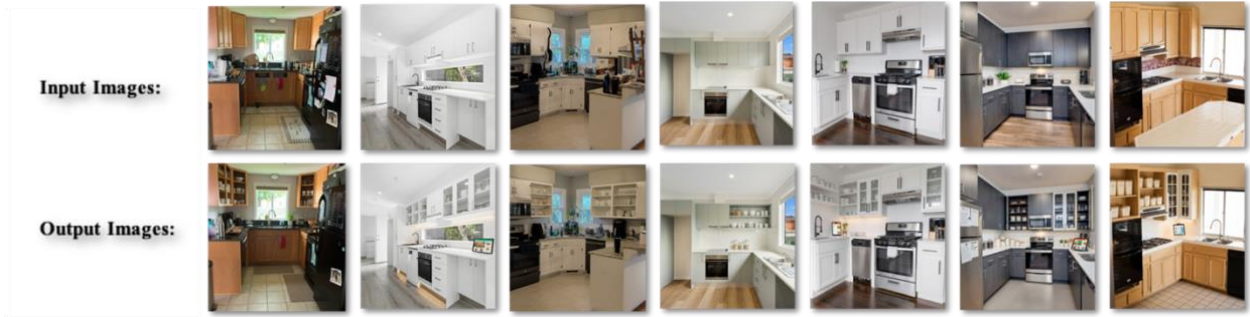

The combined technical and user evidence demonstrates that the AI system (particularly the best-performing M1 model) successfully generated MCI-friendly kitchen designs that align with HDG recommendations and are strongly preferred by the target population. These findings fully address all three research questions and establish the models as practical visual tools for accessible home modification.

## Discussion

This study demonstrated that generative AI can effectively transform ordinary kitchen photographs into MCI-friendly designs aligned with the HDG. By combining technical model evaluation with direct feedback from 33 older adults and caregivers, the findings provide strong support for all three research questions and offer a practical pathway toward more accessible home modifications.

The AI models, particularly the best-performing M1 variant, successfully generated designs that incorporated key HDG features, including open layouts, transparent cabinetry, improved lighting, and reduced visual clutter. Although normalized CLIP scores showed a modest decline from earlier to later models, this pattern indicated successful specialization rather than reduced quality. Later models deliberately diverged from generic kitchen imagery to better represent MCI-specific accessibility elements that are underrepresented in standard training data. Similarly, GIQA metrics confirmed that the models maintained acceptable levels of visual realism while learning a coherent MCI-friendly aesthetic. These technical outcomes directly affirm that stable diffusion models enhanced with DreamBooth LoRA and ControlNet can generate functionally relevant designs (RQ1 and RQ2). Because M1 provided the optimal balance across all metrics, it was used to create the six image pairs evaluated by participants.

User validation provided even stronger evidence of real-world value. Participants overwhelmingly preferred the M1-generated kitchens as more cognitively friendly (87.4% of choices), reported high confidence in their selections, and rated the visualizations as highly helpful for planning home modifications. Thematic analysis reinforced that open or glass-front shelving, reduced clutter, and enhanced lighting were the most valued features because they reduced memory demands and visual search time. These findings align closely with Bilau et al. (2025), who showed that visible storage measurably lowers cognitive load and improves task efficiency during meal preparation for people with MCI. The secondary theme of practical realism also echoes previous work highlighting the need for balanced designs that remain maintainable in daily life (Johansson et al., 2011).

The results have important practical implications. By enabling individuals and families to upload a photo of their existing kitchen and instantly see evidence-based modifications, this AI approach removes major barriers that previously limited the adoption of the HDG. This is especially valuable for lower-income households and those without access to professional designers. The high perceived helpfulness ratings suggest that such tools can empower older adults and caregivers to make simple do-it-yourself (DIY) changes that enhance safety, independence, and quality of life. These outcomes support the broader goal of aging in place and build resilience in the face of cognitive changes.

The study also contributes to the emerging literature on generative AI in accessible design. While previous work has explored AI for general residential layouts (Zhou & Pan, 2025), this research is among the first to combine technical development with direct validation from the MCI community. The strong user preference and clear thematic support extend earlier findings on environmental modifications that reduce fall risk and support IADLs (Dalvand et al., 2024; Gitlin et al., 2006; Jekel et al., 2015).

Overall, the integration of technical performance and user-centered validation demonstrates that AI-generated visualizations offer a scalable, low-cost solution to a long-standing challenge in cognitive accessibility. This approach bridges the gap between evidence-based guidelines and practical implementation, with meaningful potential to improve daily living for older adults with MCI and their care partners.

**Limitations and Future Work**

Although this study produced promising results, several limitations should be noted. The training dataset of approximately 100 images from Unsplash was modest and may not fully capture the diversity of real-world kitchens or cultural preferences in design. The technical metrics (CLIP and GIQA) are useful for semantic and visual assessment but do not measure actual functional performance in everyday use. In addition, while the survey with 33 participants from the Emory cognitive empowerment program provided valuable user validation, it relied on static image pairs rather than in-home trials or long-term experience with the modifications. The convenience sample was also predominantly U.S.-based and may limit generalizability across broader populations.

Future work will address these gaps in several ways. First, the training dataset will be expanded with more diverse kitchen images and transfer learning from larger architectural collections to improve cultural and stylistic coverage. Second, a user-friendly web interface can be developed so individuals with MCI and their caregivers can upload photos of their own kitchens and generate personalized modifications instantly. Third, longitudinal studies will track actual home changes and measure outcomes such as safety, independence, and caregiver burden. Finally, the approach will be extended to other key living spaces, including bathrooms and entryways, to create a more comprehensive solution for aging in place.

## Conclusion

This study demonstrates the power of generative artificial intelligence to translate the HDG into clear, actionable visual tools for people living with MCI. The models, especially the best-performing M1 version, successfully produced designs featuring open layouts, transparent cabinetry, improved lighting, and reduced clutter. Technical metrics confirmed strong alignment with HDG recommendations, while the survey of 33 older adults and caregivers showed overwhelming preference for the AI-modified kitchens (87.4%) and high ratings for helpfulness in planning modifications. Participants particularly valued the increased visibility and reduced cognitive load provided by open shelving and visual aids.

These findings directly address the research questions and offer a practical, low-cost solution that empowers families to visualize and implement simple DIY changes without professional assistance. By making evidence-based modifications easier to understand and adopt, this AI approach can enhance safety, independence, and quality of life for older adults with MCI. As the population ages, such accessible technologies will play a vital role in supporting cognitive health, reducing caregiver burden, and promoting resilient communities where people can age in place with dignity.

## Acknowledgements

We thank Bolaji Omofojoye for her invaluable contributions to this study. We also thank the members, care partners, and staff at the Charlie and Harriet Shaffer Cognitive Empowerment Program and the Emory Brain Health Center for participating in this study and supporting our research.

## Declaration of Generative AI and AI-Assisted Technologies in the Writing Process

The author declares that Grammarly, an AI-assisted writing software, was used in proofreading and refining the language used in the manuscript. The usage was limited to correcting grammatical and spelling errors and rephrasing statements for accuracy and clarity. The author also declares that Claude AI, a generative AI platform, was used in generating Figure 1 using content generated by the author. The author further declares that, apart from Grammarly and Claude AI, no other AI or AI-assisted technologies have been used to generate content in writing the manuscript. The ideas, design, procedures, findings, analyses, and discussion are originally written and derived from careful and systematic conduct of the research.

**Contact email(s):** bilauibrahim@gmail.com, eunhwa.yang@design.gatech.edu, ibilau3@gatech.edu

## Appendix A

# AI Kitchen Design Survey – MCI

## Informed Consent

You are being asked to be a volunteer in a research study. The purpose of this study is to explore how artificial intelligence (AI) can generate home design images, particularly kitchens that support the needs of individuals with Mild Cognitive Impairment (MCI). By understanding how people perceive and evaluate different AI-generated kitchen layouts in terms of preference and realism, your responses will help refine the AI model to create more accessible and user-friendly designs.

The online survey will take approximately 10–15 minutes to complete. Responses will be kept confidential, anonymous (no identifiers collected), and used solely for academic research reported in aggregate form; we will comply with any applicable laws and regulations regarding confidentiality. The risks involved are no greater than those involved in daily activities. You will not benefit or be compensated for joining this study.

To make sure that this research is being carried out in the proper way, the Georgia Institute of Technology IRB may review study records. The Office of Human Research Protections may also look at study records. If you have any questions about the study, you may contact Ibrahim Bilau at (850) 345-3488 or ibilau3@gatech.edu. If you have any questions about your rights as a research subject, you may contact Georgia Institute of Technology Office of Research Integrity Assurance at IRB@gatech.edu.

Thank you for participating in this study.

**Consent Confirmation: Please confirm the following before proceeding.**
- ○ I voluntarily agree to participate in this study.
- ○ I understand I can withdraw at any time without penalty.

*By completing the online survey, you indicate your consent to be in the study. Your participation is entirely voluntary, and you may choose to skip any question or withdraw from the survey at any time.*

## Kitchen Comparisons

**Q1** Please look at the two kitchens below. Which kitchen looks more cognitively friendly, i.e., easier to understand, safer, and easier to use for memory or attention difficulties?

| **Kitchen Image A** | **Kitchen Image B** |
|---|---|
| [Image A] | [Image B] |

- ○ Kitchen Image A
- ○ Kitchen Image B

**Q2** How confident are you in your choice of the kitchen Image above?

| 1 = Not confident at all | 2 = Slightly confident | 3 = Somewhat confident | 4 = Moderately confident | 5 = Mostly confident | 6 = Very confident | 7 = Extremely confident |
|---|---|---|---|---|---|---|
| ○ | ○ | ○ | ○ | ○ | ○ | ○ |

**Q3** In one sentence, what made the kitchen you chose easier or safer to use for memory or attention difficulties?

[ ]

**Q4** Which kitchen looks more like a real home kitchen you might implement?
- ○ Image A looks much more realistic
- ○ Both look equally realistic
- ○ Image B looks much more realistic

## Post Comparison

**Q5** How helpful do you think AI-generated kitchen visualizations would be for planning home modifications?

| 1 = Not at all helpful | 2 = Not helpful | 3 = Slightly unhelpful | 4 = Neutral | 5 = Slightly helpful | 6 = Helpful | 7 = Extremely helpful |
|---|---|---|---|---|---|---|
| ○ | ○ | ○ | ○ | ○ | ○ | ○ |

**Q6** Please briefly explain any general thoughts about the kitchens or features that made them easier or harder to use.

[ ]

## Familiarity and Demographics

**Q7** How often do you cook or help in the kitchen?
- ○ Every day
- ○ A few times a week
- ○ Occasionally
- ○ Rarely

**Q8** What is your age group?
- ○ 50–64
- ○ 65–74
- ○ 75+

**Q9** What is your relationship to MCI?
- ○ Older adult with MCI

- Older adult without MCI
- Caregiver/partner
- Other

## End of Survey

Thank you for participating! Your responses will help improve the design of safer, more accessible kitchens for older adults and caregivers.